\documentclass[a4paper,12pt]{article}
\usepackage{fullpage}
\usepackage{amssymb,amsmath}
\usepackage{xspace}
\usepackage{graphicx}
\usepackage{amssymb}
\usepackage{amsmath}
\usepackage{multicol}
\usepackage{rotating}
\usepackage{caption}
\usepackage{subcaption}
\usepackage{float}
\usepackage{hyperref}

\newcommand{\pa}{\partial}
\newcommand{\pr}{\partial}

\newcommand{\hog}{Hedgehog\xspace}
\newcommand{\sk}{Skyrmion\xspace}
\newcommand{\sks}{Skyrmions\xspace}
\newcommand{\bx}{\mbox{\boldmath $x$}}

\newcommand{\fgs}{figures\xspace}

\newcommand{\btau}{\mbox{\boldmath $\tau$}}
\newcommand{\bpi}{\mbox{\boldmath $\pi$}}
\newcommand{\bv}{\mbox{\boldmath $v$}}
\newcommand{\bphi}{\mbox{\boldmath $\phi$}}
\begin{document}

\title{\vskip -70pt
\vskip 50pt
{\bf \Large \bf The decay of Hopf solitons in the Skyrme model}
 \vskip 30pt
\author{
David Foster$^\dagger$ \\[40pt]
{\em \normalsize $^\dagger$ School of Physics, HH Wills Physics 
Laboratory,}\\[0pt] 
{\em \normalsize University of Bristol, Tyndall Avenue, 
Bristol BS8 1TL, U.K.}\\ 
{\normalsize Email: \quad dave.foster@bristol.ac.uk}\\[5pt]
}}

\maketitle
\vskip 20pt

\maketitle
\begin{abstract}
It is understood that the Skyrme model has a topologically interesting baryonic excitation which can model nuclei. So far no stable knotted solutions, of the Skyrme model, have been found. Here we investigate the dynamics of Hopf solitons decaying to the vacuum solution in the Skyrme model. In doing this we develop a matrix-free numerical method to identify the minimum eigenvalue of the Hessian of the corresponding energy functional. We also show that as the Hopf solitons decay, they emit a cloud of isospinning radiation.

\end{abstract}
\section{Introduction}

The Skyrme model \cite{Skyrme:1961vq} is a nonlinear theory of pions which was identified by Witten as a low energy 
effective model of QCD \cite{Witten:1983tw, Witten:1983tx}.
The model has a conserved 
topological charge that is interpreted as the baryon number $B$, and the minimal 
energy static solutions for each integer $B$ are topological soliton solutions called \sks. 

No known stable knotted solutions have been found in the classical Skyrme model. This is because, unlike the Skyrme-Faddeev model \cite{Faddeev1975,Sutcliffe:2007ui}, the model does not possess the necessary structure to stabilise such configurations. Here we investigate how Hopf solitons dynamically decay into the vacuum solution. This analysis requires developing an understanding of the geometry, about a point, of configuration space. Similar analysis has recently been performed for the Skyrme model, namely understanding the vibrational modes of the Skyrmions which correspond to Lithium-7 \cite{Halcrow:2015rvz} and Oxygen-16 \cite{Halcrow:2016spb}.
The format of this paper is as follows: We introduce the Skyrme mode. We then numerically show that the first seven Hopf solitons are solutions of Skyrme model, but not minimum energy solutions. Next we discuss how the Hopf solitons decay into the vacuum solution, where we introduce a matrix-free numerical method to identify the prominent direction of breakup. At the end we discuss how the Hopf solitons emit a cloud of isospinning radiation, followed by a short conclusion.

\section{The Skyrme model}
The Skyrme field, $U(\bx)$, is an $\mbox{SU}(2)$-valued scalar and can be expressed as $U(t,\bx)=\sigma(t,\bx) \, I_2 + i \bpi(t,\bx)\cdot \btau$, where $\sigma(t,\bx)$ is a constraint field, $\pi_1(t,\bx),\pi_2(t,\bx),\pi_3(t,\bx)$ are the three pion fields, $\btau=(\tau_1,\tau_2,\tau_3)$ are the three Pauli matrices and the constraint $\sigma^2+\bpi\cdot\bpi =1$.
Here it is convenient to represent the Skyrme field as a four component unit vector $\bphi(t,\bx)=(\sigma(t,\bx),\pi_1(t,\bx),\pi_2(t,\bx),\pi_3(t,\bx))$, where $\bphi(t,\bx)\cdot\bphi(t,\bx)=1$. The model is Lorentz invariant and, in so-called Skyrme units, it can be defined by the Lagrangian density,

\begin{align} \label{Ls}
\mathcal{L}=\pa_\mu \bphi \cdot \pa^\mu \bphi -\frac{1}{2}(\pa_\mu \bphi \cdot\pa^\mu \bphi)^2+\frac{1}{2}(\pa_\mu \bphi \cdot\pa_\nu \bphi)(\pa^\mu \bphi \cdot\pa^\nu \bphi).
\end{align}
The static energy functional associated with this Lagrangian density is,
\begin{align}
E= \frac{1}{12\pi^2} \int \left( \pr_i \bphi \cdot \pr_i \bphi +\frac{1}{2}(\pr_i \bphi \cdot\pr_i \bphi)^2-\frac{1}{2}(\pr_i \bphi \cdot\pr_j \bphi)^2 \right)d^3 x. \label{Skyrme-eng}
\end{align}

At fixed time, $\bphi(\bx)$ is a map $\bphi:\mathbb{R}^3\to
S^3$, where $\bphi \to (0,0,0,1)$ at spatial infinity. This boundary
condition compactifies $\mathbb{R}^3\cup\{\infty\}$ to $S^3$. Hence, a finite energy
configuration $\bphi(\bx)$ extends to a map $\bphi(\bx):S^3 \to S^3$, and
then belongs to a class of $\pi_3(S^3) = \mathbb{Z}$ and therefore is indexed by an 
integer $B \in \mathbb{Z}$, called the baryon number. $B$ is also 
the degree of the map $\bphi(\bx)$ which can be explicitly calculated as
\begin{equation}
B \equiv \int b(\bx) \, d^3x = \frac{1}{2\pi^2} \int
\varepsilon_{abcd}\bphi_a \pa_1\bphi_b\pa_2\bphi_c\pa_3\bphi_d \, d^3x \,, \label{baryon-density}
\end{equation}
where $b(\bx)$ is the baryon density. Static \sks are solutions of the equations of motion, which are derived from variation of energy, $\delta E(\bphi_s)=0$. They are the minimal energy solutions for each value of 
$B$. In the \fgs of Skyrmions we plot level-sets of baryon density $b(\bx)$.

\begin{figure}[H] 
       \centering
       \begin{subfigure}[b]{0.3\textwidth}
               \centering
    \includegraphics[width=\textwidth]{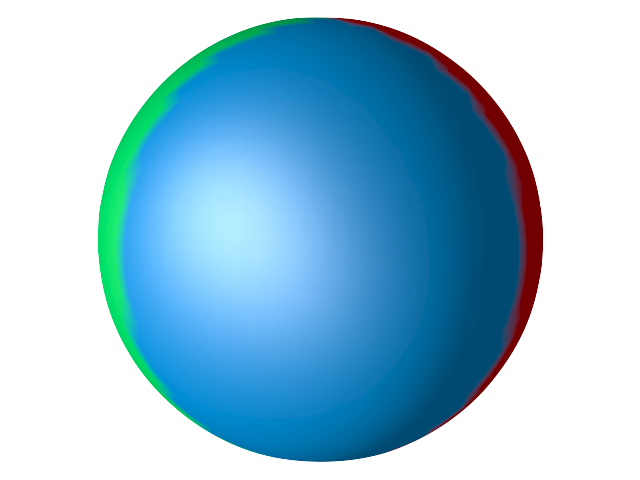}
               \caption*{Side}
    
       \end{subfigure}%
        ~ %add desired spacing between images, e. g. ~, \quad, \qquad etc. 
        \begin{subfigure}[b]{0.3\textwidth}
                \centering \includegraphics[width=\textwidth]{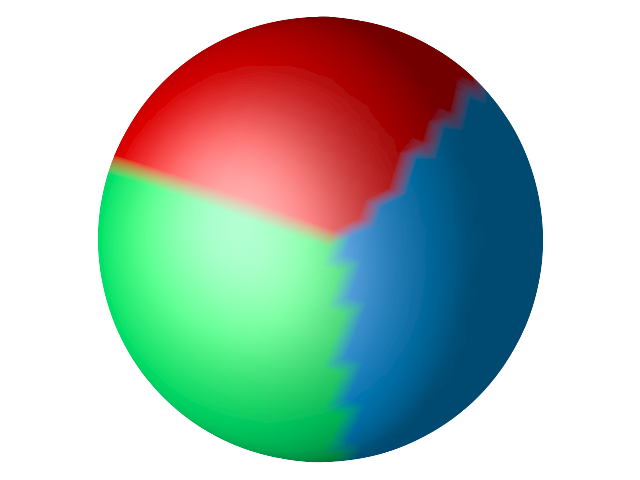}
                \caption*{Top}
         
        \end{subfigure}
        \caption{$B=1$ \hog \sk.}
        \label{B1Skyrme}
\end{figure}

Figure \ref{B1Skyrme} shows a $B=1$ \hog \sk coloured as
in \cite{Manton:2011mi}. 
The centres of the red, green and blue regions are where $\pi_3=0$ and
$\tan^{-1}\left(\frac{\pi_2}{\pi_1}\right)=0,\frac{2\pi}{3},\frac{4\pi}{3}$
respectively. The unseen white region is where $\pi_3=1$ (the boundary) and the black is where $\pi_3=-1$ (centre of the Skyrmion). This is the pion colouring scheme used throughout this paper.

The Skyrme-Faddeev model \cite{Faddeev1975, Sutcliffe:2007ui} is similar to the Skyrme model, but the field maps to the 2-sphere not the 3-sphere. Such a map belongs to a homotopy class of $\pi_3(S^2)= \mathbb{Z}$. Configurations, plotted as preimages of a point on the target two sphere, in $\mathbb{R}^3$ are knotted string-like solitons \cite{Sutcliffe:2007ui}. Importantly, if we set $\sigma(t,\bx)=0$ in equation \eqref{Ls}, we gain the Skyrme-Faddeev model (with a different normalisation, multiplying the solutions in \cite{Sutcliffe:2007ui} by $\frac{8\sqrt{2}}{3}$ yields the energy in the chosen Skyrme units). In this case the field $\bphi$ maps to an equatorial sub unit 2-sphere $S^2\subset S^3$. The Skyrme-Faddeev minimum energy solutions have been known for a long time \cite{Sutcliffe:2007ui,Foster:2010zb}. Here we are interested in Hopfion like solutions in the Skyrme model.  A Skyrmion is a topological soliton, where the image of the map is the entire 3-sphere and the image
for a Hopfion, in the Skyrme-Faddeev model, is the 2-sphere. In both cases the images are not homotopic to a point, and the solitons are stabilised for all continuous deformations. Unlike Skyrmions, and the conventional Hopfions, the Hopfions in the Skyrme model are not energetic minima in each topological sector. The image of the field does not cover the entire 3-sphere. We call this degree quasi-topology and we label the degree of each configuration as $Q'\in\mathbb{Z}$. The Hopfion-like solutions are called quasi hopfions.

\section{Numerical solutions}
Before studying the decay of quasi hopfions in the Skyrme model, we need to attain static quasi hopfion solutions in the Skyrme model. Taken from the Skyrme-Faddeev model, \cite{Sutcliffe:2007ui,Foster:2010zb}, initial conditions for $Q'=1-7$ are energetically relaxed on a lattice of $100\times100\times100$ with spacing $\delta x=0.1$, with derivatives approximated by fourth-order accurate finite differences. This produces solutions, $\bphi_s(\bx)$, which are extrema of the energy functional. These extrema are found by solving the equations of motion, namely the variation of the energy functional, $\delta E(\bphi_s)=0$.
\begin{figure}[H]
       \centering
    \includegraphics[width=0.75\textwidth]{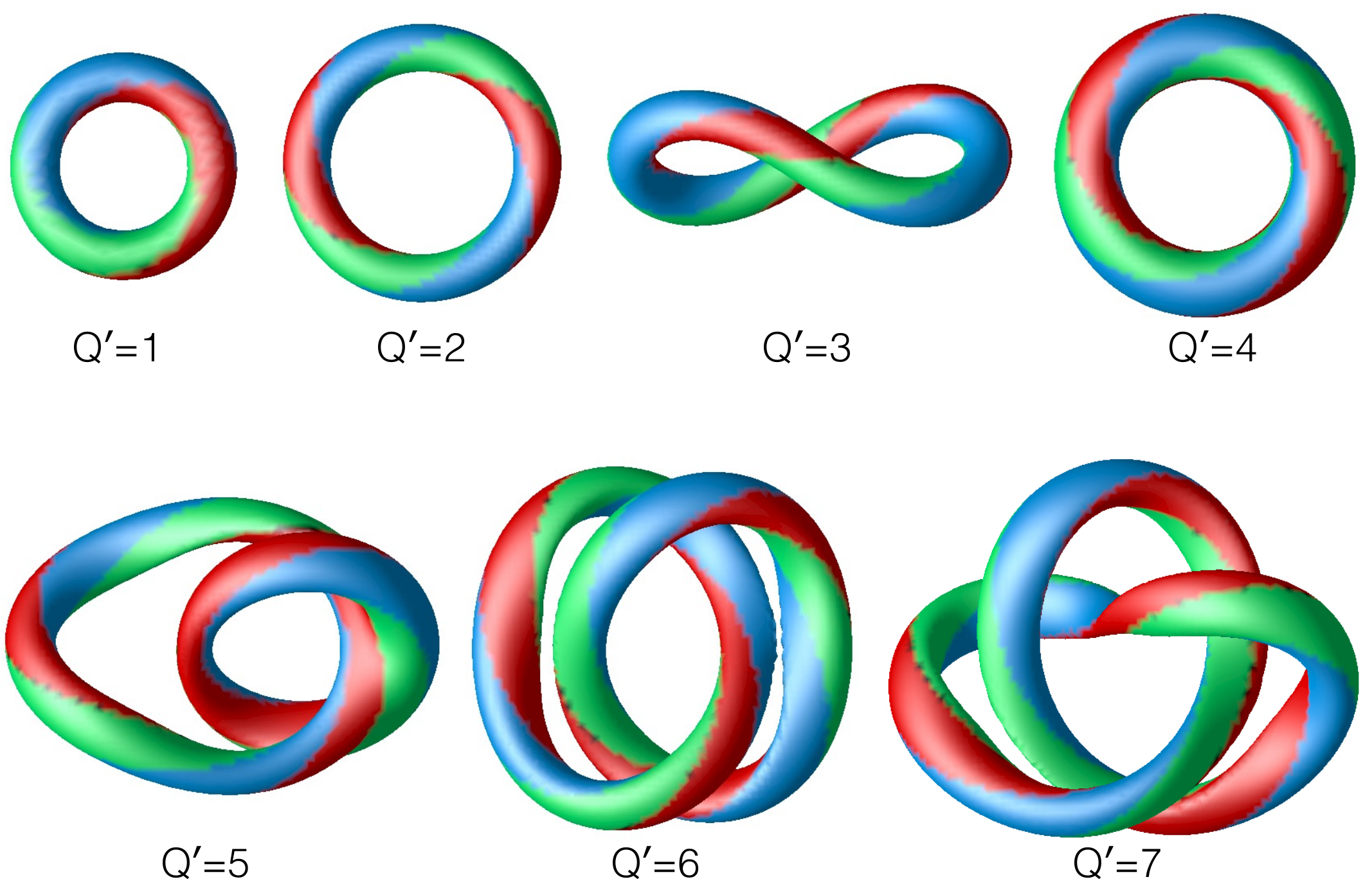}
               \caption{Hopfions in the Skyrme model, tubes are level sets of $\pi_3 \approx -1$ and colouring is due to the pion fields as in the text.}
        \label{knots}
\end{figure}
All of these solutions are shown in figure \ref{knots}, it must be noted that these are not global minima. This is because they can all be continuously deformed into the vacuum solution, which classically has zero energy. These solutions are the same as those found in the Skyrme-Faddeev model \cite{Sutcliffe:2007ui}. The solutions verify that the quasi hopfions are indeed solutions of the Skyrme model's equation of motion, stabilised by symmetry.

\section{Breakup modes}
Y.M. Cho, B.S. Park and P.M. Zhang \cite{cho2008new}, used an energetic argument to conjecture that the $Q'=1$ quasi hopfion would breakup into a Skyrmion anti-Skyrmion pair (baryon anti-baryon pair). It is correct that the $Q'=1$ quasi hopfion, with energy $E=4.54$ in Skyrme units, has more energy than an infinity separated Skyrmion anti-Skyrmion which as energy $E=2.464$. However, it is not immediately apparent that quasi hopfions would decay in a direction in configuration space, $\mathcal{C}$, such that the domain would cover the target space. Where the configuration space $\mathcal{C}$ is defined as,

\begin{align}
\mathcal{C}:=\{ \bphi:\mathbb{R}^3\to S^3 \subset \mathbb{R}^4 | \bphi \to (0,0,0,1) ~\mbox{as}~ |\bx|\to\infty \} \nonumber.
\end{align}

\begin{figure}[H]         
                \centering \includegraphics[width=\textwidth]{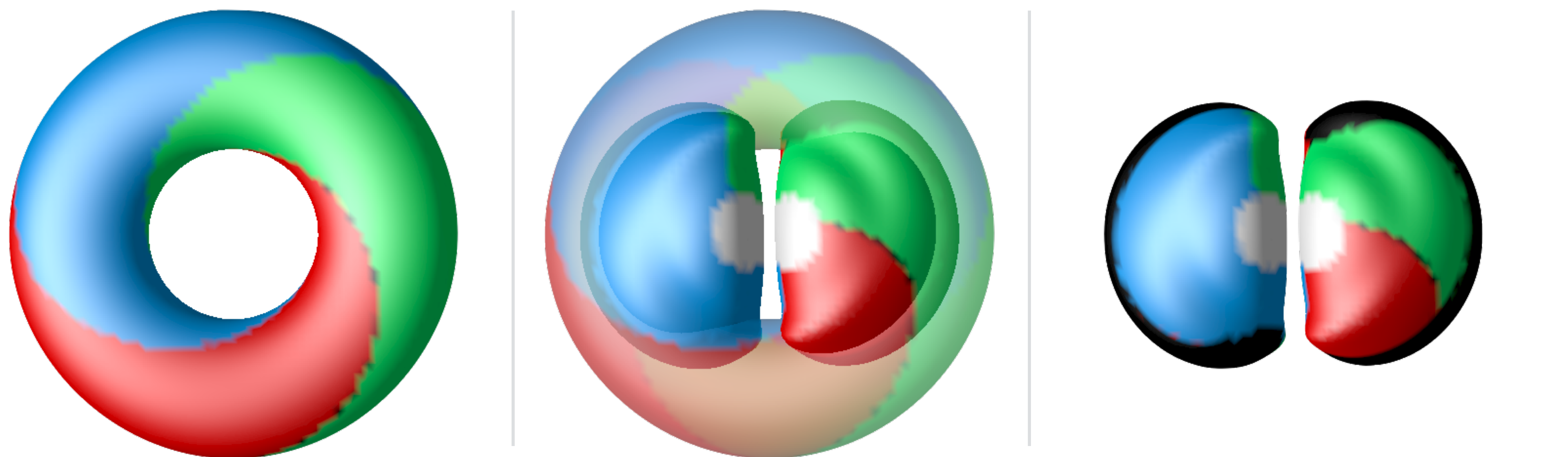}
         \caption{Field configuration of a simulation with initial condition $\bphi=\bphi_s+\bv_a$ when the Skyrmion anti-Skyrmion is formed. The left image show a level set of $\pi_3\approx-1$ (hopfion). The central image show the hopfion and the Skyrmion anti-Skyrmion in the centre of the quasi hopfion. The right most image shows the Skyrmion and anti-Skyrmion in isolation. All of the colours indicate the pion fields.}
         \label{perturbation-arb-Q1}
\end{figure}

Another way to understand this is to consider the quasi hopfion configurations as an equatorial two sphere on the 3-sphere. Such a 2-sphere can be continuously shrunk moving on the 3-sphere, until it contracts to the point which represents the vacuum solution. This homotopy would be such that every preimage has baryon density (Jacobi determinant) zero, and hence no region of Skyrmion or anti-Skyrmion.\\
To test how these quasi hopfions break up we first add a small arbitrary periodic perturbation, $\bv_a(\bx)=(0.0001\sin(x),0,0,0)$, to a solution $\bphi(\bx)=\bphi_s(\bx)+\bv_a(\bx)$. This configuration is then normalised such that $\bphi(\bx)\cdot\bphi(\bx)=1$. Using this as an initial condition, in a full time evolution simulation, produces the images in figure \ref{perturbation-arb-Q1}.

This simulation flows to a configuration with $\int |b(\bx)| d^3x\approx 2$ and then breaks down. This implies that the quasi hopfion does break up into a Skyrmion anti-Skyrmion pair, which then annihilate. Also, the colouring in figure \ref{perturbation-arb-Q1} show that the Skyrmion anti-Skyrmion pair cover the entire target 3-sphere. Fundamentally this does not show that the quasi hopfion generically breaks up into a Skyrmion anti-Skyrmion pair.

To understand how these quasi hopfions decay, we need to understand the directions of greatest reduction in energy about a point in configuration space. We are interested in critical points in the configuration space which correspond to solutions of the equations of motion. More particularly, points $\bphi_s(\bx)$ such that $\delta E(\bphi_s(\bx))=0$; namely solutions of the equations of motion of $E$. To find the prominent break-up perturbation $\bphi=\bphi_s+\bv$ we need to find $\bv$ in $\mathcal{C}$ which is the greatest reduction in energy about $\bphi_s$. To do this we linearise the energy at $\bphi_s$ as $E(\bphi)=E(\bphi_s)+\bv^\intercal \nabla E(\bphi_s)+ \frac{1}{2}\bv^\intercal H_E(\bphi_s) \bv +...$, where $H_E$ is the Hessian of $E(\bphi_s)$. We must include at least second order terms because, by the fundamental theorem of calculus, $\bv^\intercal \nabla E(\bphi_s)=0$ for all $\bv$. The usual procedure is to find the smallest eigen-pair  of the Hessian, $H_E$. To do this numerically on a lattice of $100^3$ lattice the Hessian matrix would be a $4$ million by $4$ million matrix. Finding the eigenvalues of such a large (partially sparse) matrix is limited by the memory of most computers. Populating such a matrix is also very involved. To proceed we make use of the fact that the second directional derivative of $E$ in $\bv$ is $D^2_{\bv} E(\bphi)= \bv^\intercal H_E(\bphi) \bv$. So finding $\bv$ which minimises $D^2_{\bv}E(\bphi_s)$ will produce the optimal perturbation. To find the $\bv$ which minimises $D^2_{\bv} E(\bphi_s)$ we implemented a gradient flow algorithm,
\begin{align}
\dot{\bv}&= - \frac{\delta}{\delta \bv}D^2_{\bv} E(\bphi_s). \\ \nonumber
\end{align}
It should be noted that this procedure is the same as linear stability, where one takes the equations of motion (first order directional derivative), linearises them to give the Hessian, $H_E$, (second order directional derivative). One can then find the smallest eigenvalue $\lambda$ by rearranging the eigenequation to give $\lambda=\frac{\bv^\intercal H_E\bv}{\bv^\intercal \bv}$, which can be minimised by finding the $\bv$ which minimises $\bv^\intercal H_E\bv$. As shown previously, this is the same as finding the $\bv$ which minimises $D^2_{\bv}(\bphi_s)$.  The main advantage of this method, over the explicit Hessian matrix-type approach, is that it does not require an explicit formulation of the Hessian matrix, it is known as a matrix-free method. Along with the gradient flow we include the constraints $(\bphi_s+\bv)\cdot(\bphi_s+\bv)=1$ and $\max(\bv)<\mbox{const}$, these conditions stop $\bv$ from having a large component pointing towards the centre of the target $S^3\subset\mathbb{R}^4$.

\begin{figure}[H]         
                \centering \includegraphics[width=\textwidth]{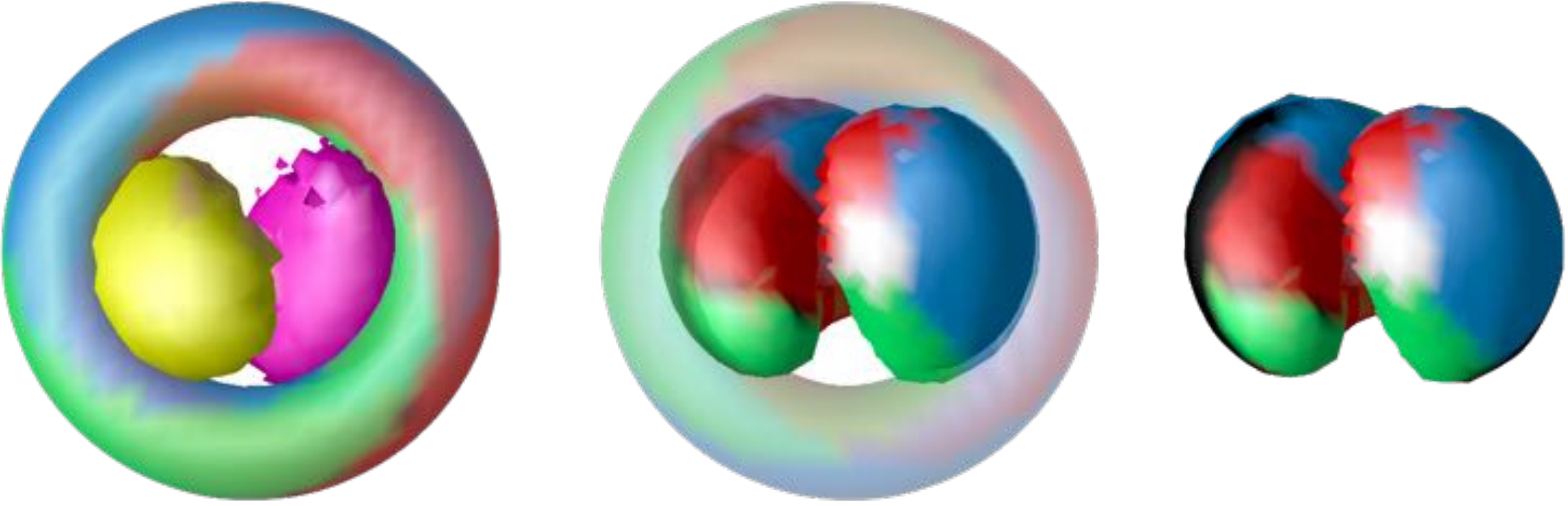}
         \caption{Optimal perturbation break-up, $\bphi=\bphi_s+\bv$, for the $Q'=1$ quasi hopfions. The left image is of the quasi hopfion, the purple and yellow lumps are positive and negative baryon density respectively. The central image shows the quasi hopfion and level sets of positive and negative baryon density. The right images shows level sets of positive and negative level sets of baryon density. Red, blue, green, white and black colouring indicates the pion fields.}
         \label{perturbation-Q1}
\end{figure}

Performing this gradient flow indeed showed that the preferred break-up direction is for the quasi hopfion to decay into what seems like a Skyrmion anti-Skyrmion pair, but aligned differently to the quasi hopfion, as shown in figure \ref{perturbation-Q1}.

%\begin{figure}[H]         
%                \centering \includegraphics[width=\textwidth]{{"/Users/Dave/Documents/Dave's talks/Poland-2016/Skyrme-field-plus-vector"}.pdf}
%         \caption{Field configuration $\bphi=\bphi_s+\bv$. Level set of $\pi_3$, negative baryon density orange and positive purple.}
%         \label{perturbation-Q1}
%\end{figure}
\begin{figure}[H]         
                \centering \includegraphics[width=\textwidth]{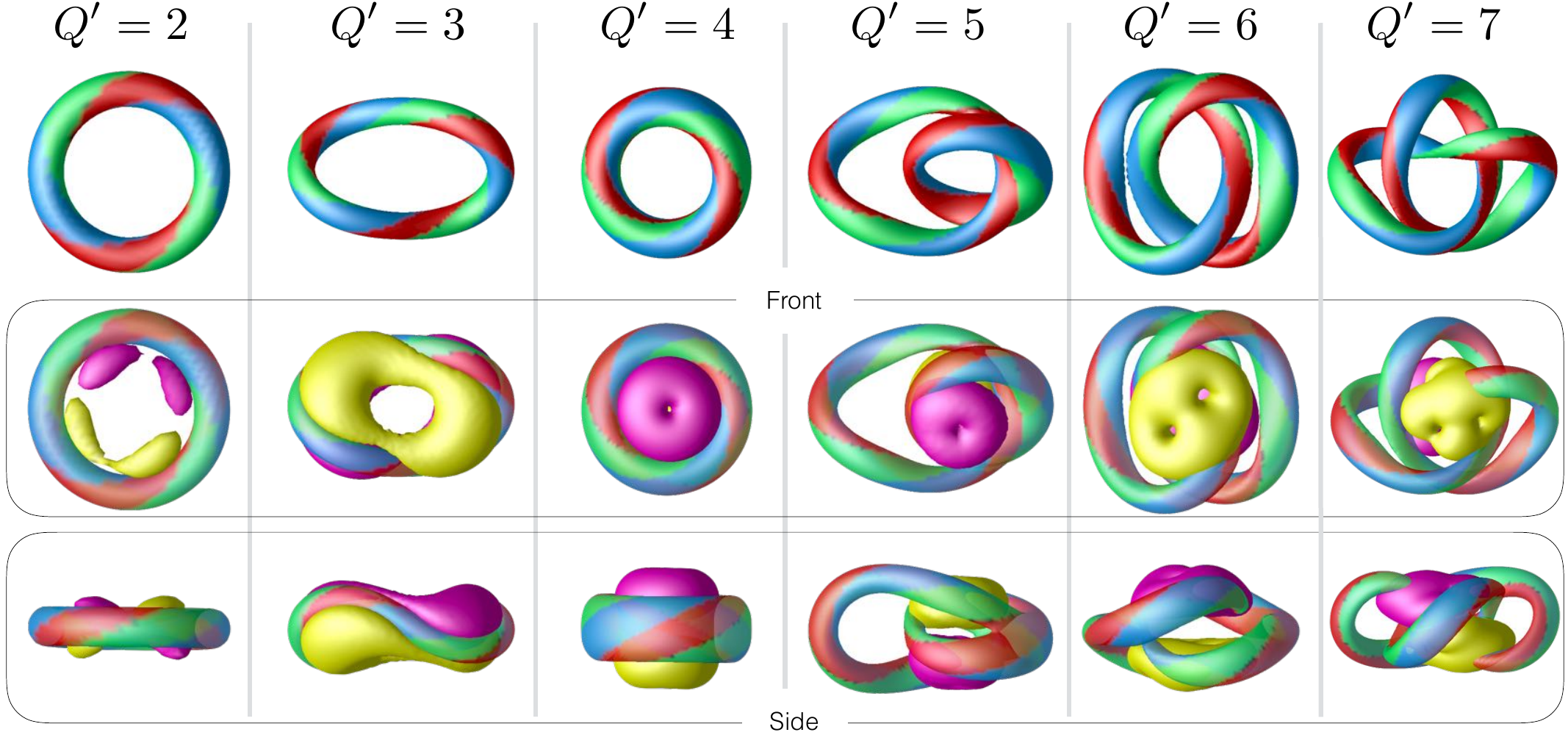}
         \caption{Optimal perturbation break-up, $\bphi=\bphi_s+\bv$, for the $1 < Q' \leq 7$ quasi hopfions. Purple and yellow lumps are positive and negative baryon density Skyrmions respectively.}
         \label{perturbation-Q2-7}
\end{figure}

Considering the Hessian's eigenequation $H_E\bv=\lambda\bx$, it is apparent that $\bv$ and $-\bv$ have the same eigenvalue. Also, the $Q'=1$ quasi hopfion is invariant under a combined isorotation, $I$, and spacial rotation, $D$, $\bphi_s(D\bx)=I\bphi_s(\bx)$. So, considering an isorotation and spacial rotation of the eigenequation, we see that $H_E(I^{-1}\bphi_s(D\bx))I^{-1}\bv(D\bx)=\lambda I^{-1}\bv(D\bx)$. This gives that $H_E(\bphi_s(\bx))I^{-1}\bv(D\bx)=\lambda I^{-1}\bv(D\bx)$, hence $\lambda$ is invariant under the combined isorotation and rotation when $\bphi_s(D\bx)=I\bphi_s(\bx)$.
The same analysis can be performed for $1 < Q' \leq 7$ and is shown in figure \ref{perturbation-Q2-7}.

\section{Imaging quasi hopfion break up}
\begin{figure}[H]         
                \centering \includegraphics[width=0.88\textwidth]{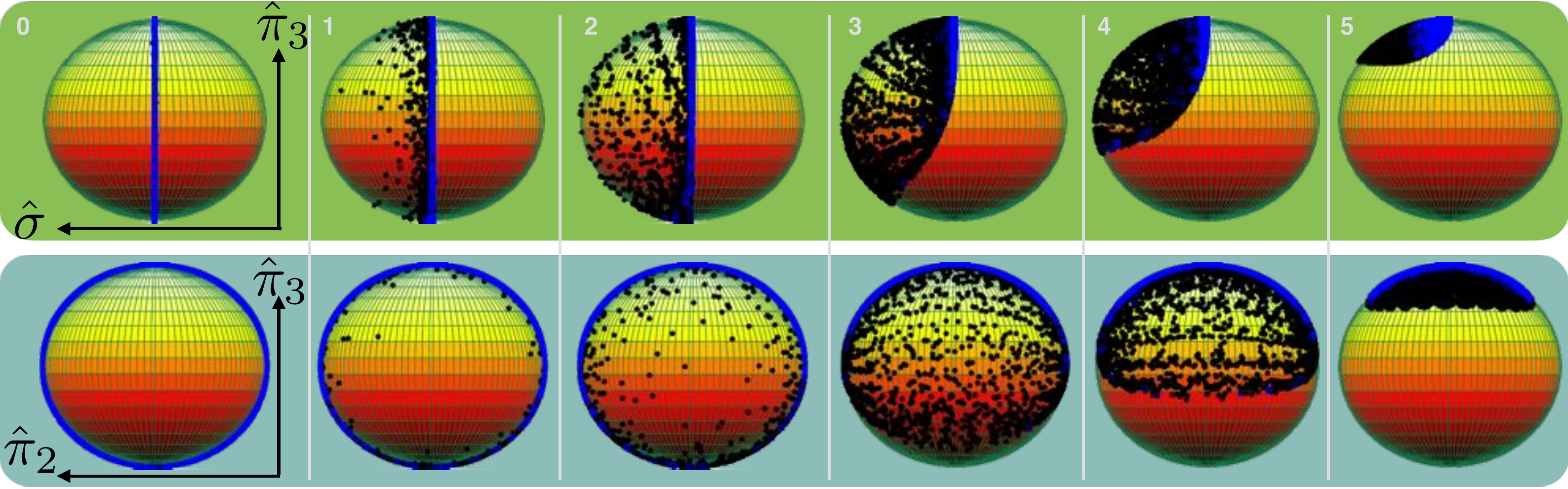}
         \caption{Optimal perturbation break-up, $\bphi=\bphi_s+\bv$, for the $Q' =1$ quasi hopfion. Points plotted on a sub sphere of the target space, black $b(\phi(\bx))\neq0$, blue $b(\phi(\bx)) =0$. At time slices chosen to show the unwrapping.}
         \label{opt-pert-Q1-target}
\end{figure}

The images in figure \ref{perturbation-arb-Q1} of the arbitrary perturbation and in figure \ref{perturbation-Q1} of the optimal perturbation are similar. However, if we use the optimal perturbation as an initial condition for a full time simulation, we find that the maximum value for $\int |b(\bx)|d^3x$ is $0.91$ and not $2$. The two decays therefore have a subtle difference. To understand the difference we can consider how the image of the quasi hopfion unwraps off the target 3-sphere. To see this, we consider the 2-sphere defined by, $\hat{\pi}_3=\pi_3/N,  \hat{\pi}_2=\pi_2/N,  \hat{\sigma}=\sigma/N,$ where $N=\sqrt{\pi_2^2+\pi_3^2+\sigma^2}$ and neglect points where $\pi_1=\pm1$. Figure \ref{opt-pert-Q1-target}  shows the points of the perturbed quasi hopfion on this 2-sphere where blue dots denote the points which have zero baryon density, $b(\bphi(\bx))=0$ (zero Jacobi determinant), and black dots denote the points which have non-zero baryon density, $b(\bphi(\bx))\neq0$ (Jacobi determinant). This, figure \ref{opt-pert-Q1-target}, shows how the quasi hopfion unwraps off the 2-sphere.

\begin{figure}[H]         
                \centering \includegraphics[width=0.88\textwidth]{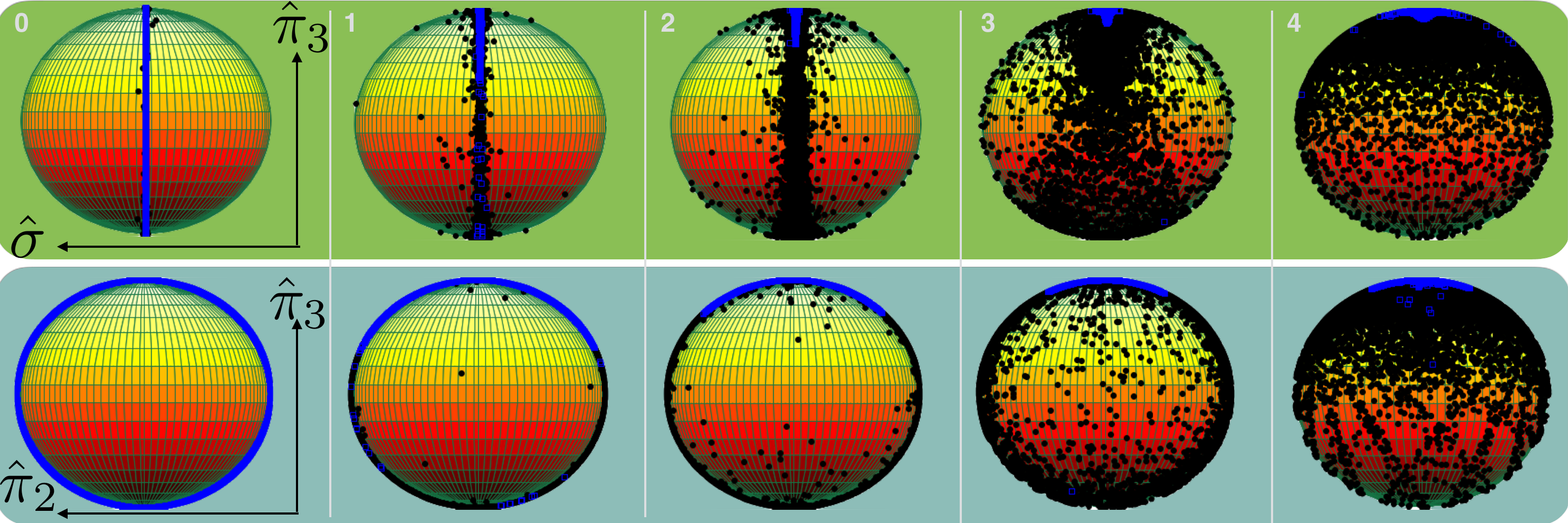}
         \caption{Arbitrary perturbation break-up, $\bphi=\bphi_s+\bv_a$, for the $Q' =1$ quasi hopfions. Points plotted on a sub sphere of the target space, black $b(\phi(\bx))\neq0$, blue $b(\phi(\bx)) =0$.  At time slices chosen to show the unwrapping.}
         \label{sim-pert-Q1-target}
\end{figure}

Figure \ref{opt-pert-Q1-target} clearly shows that under time evolution the quasi hopfion expands to cover one half of the $(\hat{\pi_1},\hat{\pi_3},\hat{\sigma})$-sphere, and then contracts to the vacuum point. It happens in such a way that the boundary conditions are preserved. Also, each black point has nonzero baryon density (Jacobi determinant) but the sum of the absolute value of the baryon density (Jacobi determinant) sums to $\approx 1$. 
Figure \ref{sim-pert-Q1-target} shows the same analysis for the previous arbitrary perturbation, $\bv_a$.
Figure \ref{sim-pert-Q1-target} shows that, for the arbitrary perturbation, the field evolves to cover the entire $(\hat{\pi_1},\hat{\pi_3},\hat{\sigma})$-sphere. It does this such that the sum of baryon density (Jacobi determinant) is zero, but the integral of the absolute baryon density equals $2$. This indicates that the two break ups are distinctly different. Yet taking the inner product of the arbitrary perturbation $\bv_a$ with the optimal perturbation $\bv$ gives a value of $\approx 0.0003$. This shows that the arbitrary perturbation has a small component in the same direction, in $\mathcal{C}$, as the optimal perturbation. Hence, they can be understood as similar perturbations, which propagate in slightly different directions towards the same minimum. The optimal perturbation is direct, while the arbitrary perturbation passes near to a Skyrmion anti-Skyrmion configuration.

\section{Isospinning quasi hopfions}
\begin{figure}[H]         
                \centering \includegraphics[width=0.5\textwidth]{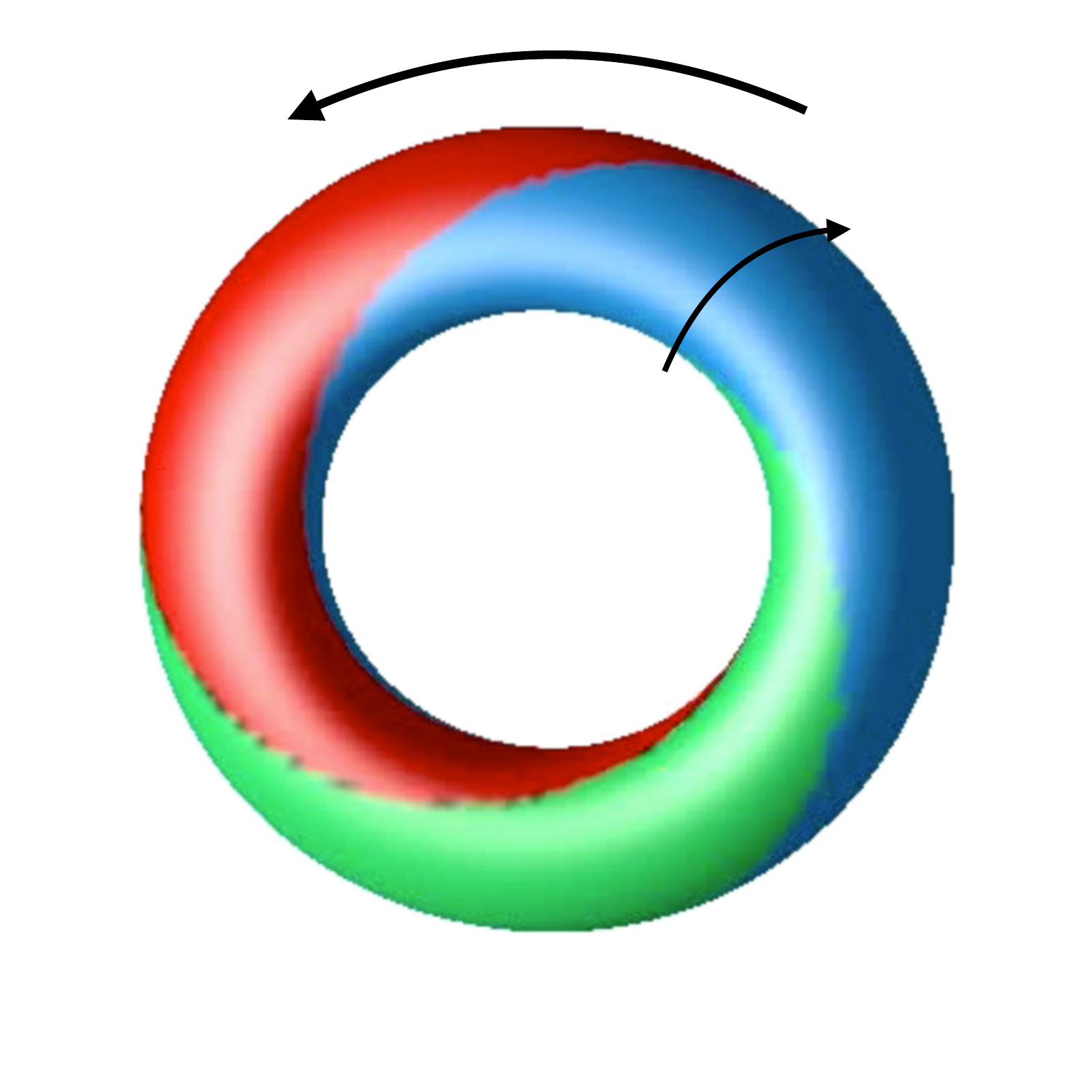}
         \caption{Isospinning $Q' =1$ quasi hopfions.}
         \label{iso-Q=1}
\end{figure}

To relate the $B=1$ Skyrmion to the proton or neutron requires including isospin. In the Skyrme model isospin acts on the pion fields, and has had some success in modelling nucleon scattering with spinning Skyrmions \cite{Foster:2015cpa}. To understand the consequence of isospinning quasi hopfions we implemented a full field simulation with a rigidly isospinning quasi hopfion as initial condition,
\begin{align}
\sigma(t,\bx)=\sigma(\bx), &~ \pi_1(t,\bx)=\pi_1(\bx)\cos(\omega  t)+\pi_2(\bx)\sin(\omega t), \nonumber \\ 
\pi_3(t,\bx)=\pi_3(\bx), &~ \pi_2(t,\bx)=\pi_2(\bx)\cos(\omega  t)-\pi_1(\bx)\sin(\omega t).
\end{align}
This is analogous to the more common $SU(2)$ formulation $U(t,\bx)=A(t)U_0(\bx)A^\dag(t)$ where $A(t)=\exp(\frac{i\omega t}{2} \tau_3)$. Such an isospin causes the $\pi_1$ and $\pi_2$ components to isospin into each other. Using this as an initial condition, with $\omega =0.28$, gives rise a configuration which is stable to perturbation for a long time. This is shown figure \ref{iso-Q=1}.

\begin{figure}[H] 
       \centering
       \begin{subfigure}[b]{0.31\textwidth}
               \centering
    \includegraphics[width=\textwidth]{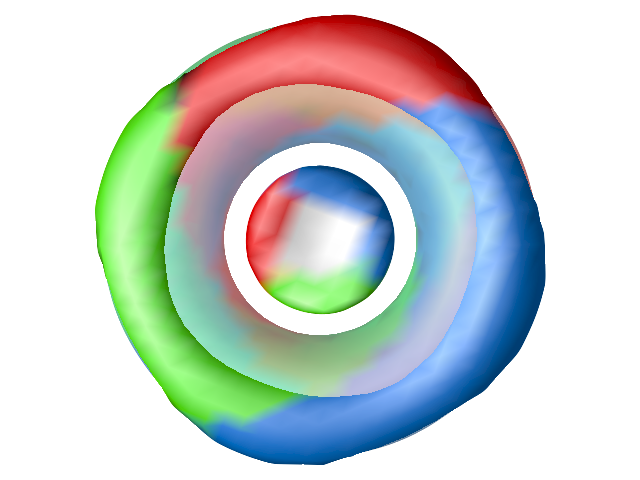}
               \caption*{$t'=-1$}
    
       \end{subfigure}%
        ~ %add desired spacing between images, e. g. ~, \quad, \qquad etc. 
        \begin{subfigure}[b]{0.31\textwidth}
                \centering \includegraphics[width=\textwidth]{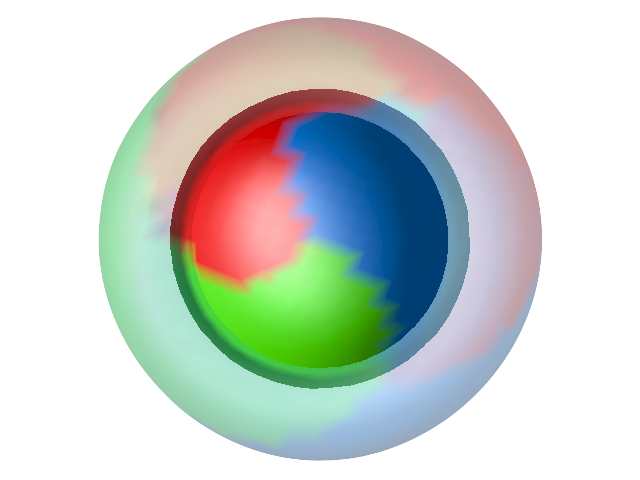}
                \caption*{$t'=0$}
                \end{subfigure}
            ~ %add desired spacing between images, e. g. ~, \quad, \qquad etc. 
        \begin{subfigure}[b]{0.31\textwidth}
                \centering \includegraphics[width=\textwidth]{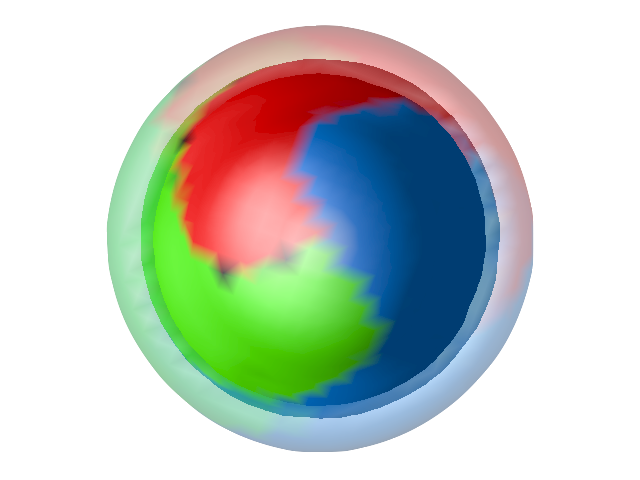}
                \caption*{$t'=1$}
        \end{subfigure}
        \caption{Transparent is the isospinning quasi hopfion, the nontransparent is a level set of pion radiation $\dot{\bphi}\cdot\dot{\bphi}$. For arbitrary time steps $t'$.}
        \label{Q1-pion-rad}
\end{figure}

The isospin, similar to the spherically symmetric $B=1$ Skyrmion, also causes a spatial rotation. This is apparent because the colours twist around the quasi hopfion, so as the colour rotate into each other they cause the ring to spin. Hence, isospin is equivalent to spin. This produces two angular momenta, one about each circumference of the quasi hopfion. As the quasi hopfion decays it radiates in the form of kinetic energy, $\dot{\bphi}\cdot\dot{\bphi}$. This kinetic energy is shown isospinning in figure \ref{Q1-pion-rad}. This cannot follow the same breakup mode as in the non isospinning case, because the Skyrmion and anti-Skyrmion would seem to spin in the same direction. This would cause a large gradient of the field in-between them, and hence a large energetic cost.

This radiation can be interpreted as massless pions being classically radiated as the isospinning hopfion decays.

\section{Conclusion}
In this work we have identified the break up of $Q'=1-7$ quasi hopfions, particularly the single $Q'=1$ quasi hopfion. We have discovered that the direction of greatest reduction in energy corresponds to the formation of a Skyrmion lump anti-Skyrmion lump pair. But there is a higher energy perturbation which gives rise to a Skyrmion anti-Skyrmion pair. We have also observed that isospinning quasi hopfions break up in to a cloud of spinning radiation, which can be understood as a cloud of isospinning classical pions. 
Mainly we have developed a matrix-free approach, based on geometry, that finds the eigenvector of the Hessian which corresponds to its minimum eigenvalue. This is a general technique, which could be applied to other field theories which have nontrivial breakup modes.
Future work would be to understand the eigenvector of the second smallest eigenvalue of the Hessian. If the second smallest eigenvalue is also negative it would show that there is also a second, higher energy, breakup perturbation.
\vspace{-0.3cm}
\section{Acknowledgments}
\vspace{-0.3cm}
I thank Prof. P.M. Sutcliffe, Dr. S. Krusch, Dr. D. Harland, Prof. M. Dennis, Dr. A. Taylor and Dr. J. Fellows for useful discussion and input. This work is funded by the Leverhulme Trust Program Grant: SPOCK, Scientific Properties Of Complex Knots.
%\bibliographystyle{/Users/Dave/Documents/Postdoc-apps/utphys}
%\bibliography{/Users/Dave/Documents/Postdoc-apps/hopf2.bib}

\end{document}